\documentclass[preprint,aps,showpacs,floats]{revtex4}
\usepackage{graphicx}
\def\lsim{\raise0.3ex\hbox{$\;<$\kern-0.75em\raise-1.1ex
\hbox{$\sim\;$}}}
\def\gsim{\raise0.3ex\hbox{$\;>$\kern-0.75em\raise-1.1ex
\hbox{$\sim\;$}}}

\begin{document}
\overfullrule 0pt
\title{ 
\vglue -2.0cm
{\small \hfill IFUSP-DFN/02-078\\
\vglue -0.4cm
\hfill IFT-P.001/2003}\\
Determining the oscillation parameters 
by Solar neutrinos and KamLAND } 
 
\author{H.~Nunokawa$^1$}\email{nunokawa@ift.unesp.br} 
\author{W.~J.~C.~Teves$^2$}\email{teves@charme.if.usp.br}  
\author{R.~Zukanovich Funchal$^2$}\email{zukanov@if.usp.br}  
 
\affiliation{\\ \\ 
$^1$ Instituto de F{\'\i}sica Te{\'o}rica,Universidade Estadual Paulista, 
     Rua Pamplona 145, 01405-900 S{\~a}o Paulo, Brazil \\ 
$^2$ Instituto de F{\'\i}sica,  Universidade de S{\~a}o Paulo  
     C.\ P.\ 66.318, 05315-970 S{\~a}o Paulo, Brazil} 
 
\begin{abstract} 
 The neutrino oscillation experiment KamLAND has provided us with the
first evidence for $\bar \nu_e$  disappearance, coming from nuclear
reactors.  We have combined their data with all solar neutrino data,
assuming two flavor neutrino mixing, and obtained allowed parameter
regions which are compatible with the so-called large mixing angle MSW
solution to the solar neutrino problem.  The allowed regions in the
plane of mixing angle and mass squared difference are now split into
two islands at 99\% C.L.  We have speculated how these two islands can
be distinguished in the near future.  We have shown that a 50\%
reduction of the error on SNO neutral-current measurement can be
important in establishing in each of these islands the true values of
these parameters lie.  We also have simulated KamLAND positron energy
spectrum after 1 year of data taking, assuming the current best fitted
values of the oscillation parameters, combined it the with current
solar neutrino data and showed how these two split islands can be
modified.
\end{abstract} 
 
 
\maketitle 
\thispagestyle{empty} 
\section{Introduction} 
\label{sec:intro} 
\vglue -0.3cm
 Many solar and atmospheric neutrino experiments have collected data
 in the last decades,  giving evidence that neutrinos produced in the Sun
 and in the Earth's atmosphere suffer flavor conversion. While the
 atmospheric neutrino results~\cite{atmnuobs} may be understood
 by $\nu_\mu \to \nu_\tau$ conversion driven by a neutrino mass
 squared difference within the experimental reach of the accelerator
 based neutrino oscillation experiment K2K~\cite{k2k}, the mass squared
 difference needed to explain the solar neutrino data was, until
 quite recently, before the Kamioka Liquid scintillator AntiNeutrino
 Detector (KamLAND)~\cite{kamland} has started its operation, too small
 to be inspected by a terrestrial neutrino oscillation experiment.

 A number of different fits, assuming standard neutrino oscillations
 induced by mass and mixing~\cite{fits} as well as other exotic flavor
 conversion mechanisms~\cite{exotics}, have been performed using the
 combined solar neutrino data from Homestake~\cite{homestake},
 GALLEX/GNO~\cite{gallex,gno}, SAGE~\cite{sage},
 Super-Kamiokande-I~\cite{superk} and SNO~\cite{sno}.  These analyses
 selected some allowed areas in the free parameter region of each
 investigated mechanism, but did not allow one to establish beyond
 reasonable doubt which is the mechanism and what are the values of the
 parameters that are responsible for solar $\nu_e$ flavor conversion.
 After the first result of the KamLAND (or KL hereafter)
 experiment~\cite{kamland} this picture has changed drastically.

 In the first part of this paper, we present the allowed region for the 
 oscillation parameters in two generations for the entire set of solar 
 neutrino data, for KamLAND data alone and for  KamLAND result combined 
 with all solar neutrino data, showing that this last result finally 
 establishes the so called large mixing angle (LMA)
 Mikheyev-Smirnov-Wolfenstein (MSW)~\cite{msw} solution 
 as the final answer to  the long standing 
 solar neutrino problem~\cite{bahcall}, 
 definitely discarding all the other mass 
 induced or more exotic solutions. 
 (For the first discussions on the complete ``MSW triangle'' which 
 includes the LMA region, see Ref.~\cite{mswtriangle}.)
 In the second part, we speculate on the possibility of further 
 constraining the oscillation parameters in the near future.
 For instance, we point out the importance of SNO neutral-current (NC)
 data in further constraining the LMA MSW solution. In particular, we discuss 
 the consequence of a significant reduction (50 \%) of the SNO 
 neutral-current data uncertainty.
 Finally, we simulate the expected inverse  $\beta$-decay $e^+$ energy 
 spectrum after 1 year of KamLAND data taking, based on the best fitted 
 values of  the oscillation parameters. We combine this with the current solar 
 neutrino data in order to show how the allowed parameter region
 can be modified.

\vglue -2.7cm
\section{Determination of Oscillation Parameters} 
\label{sec:analysis} 
\vglue -0.3cm 
 KamLAND has observed about 40\% suppression of $\bar
 \nu_e$ flux with respect to the theoretically expected
 one~\cite{kamland}, which is compatible with neutrino oscillations in
 vacuum in two generations.  In this case the relevant oscillation
 parameters, which must be determined by the fit to experimental data,
 are a mass squared difference ($\Delta m^2$) and a mixing angle
 ($\theta$).  We first obtained the allowed region in the
 ($\tan^2\theta$, $\Delta m^2$) plane compatible with all solar
 neutrino experimental data, then with KamLAND data alone, and finally
 we combine these two sets of data.  
\vglue -1.7cm

\subsection{Solar Neutrino Experiments} 
\label{subsec:solar} 
\vglue -0.3cm
 We have determined the parameter region allowed by the solar neutrino
 rates measured by Homestake~\cite{homestake}, GALLEX/GNO~\cite{gallex,gno}, 
 SAGE~\cite{sage} and SNO (elastic scattering, charged-current and 
 neutral-current reactions)~\cite{sno} (6 data points) as well as by 
 the Super-Kamiokande-I zenith spectrum data~\cite{superk} (44 data points), 
 assuming neutrino oscillations in two generations. 

 We have computed the $\nu_e \to \nu_e$  survival probability, properly taking 
 into account the neutrino production distributions in the Sun according to 
 the Standard Solar  Model~\cite{ssm}, the zenith-angle exposure of each 
 experiment, as well as the Earth matter effect as in Ref.~\cite{exotics}, 
 except that here we solved the neutrino evolution equation entirely 
 numerically. We then have estimated the allowed parameter region by 
 minimizing the $\chi^2_\odot$ function which is defined as
\begin{equation}
 \label{chi}
\chi^2_\odot = \sum_{i,j=1,...,50}
 \left[R_i^{\text{th}}-R_i^{\text{obs}} \right] \, 
 \left[\sigma_\odot^2 \right]^{-1}_{ij} \,
 \left[R_j^{\text{th}}-R_j^{\text{obs}} \right]\,,
\end{equation}
 where $R_i^{\text{th}}$ and $R_i^{\text{obs}}$ denote the
 theoretically expected and observed event rates, respectively,  
 which run through all 50 data points mentioned above, 
 and $\sigma_\odot$ is the $50\times 50$ correlated error matrix, 
 defined in a similar way as in Ref.~\cite{exotics}.
 In this work we have treated the $^8$B neutrino flux as a free parameter. 

\begin{figure} 
\centering\leavevmode 
\hglue -0.2cm
\includegraphics[scale=0.63]{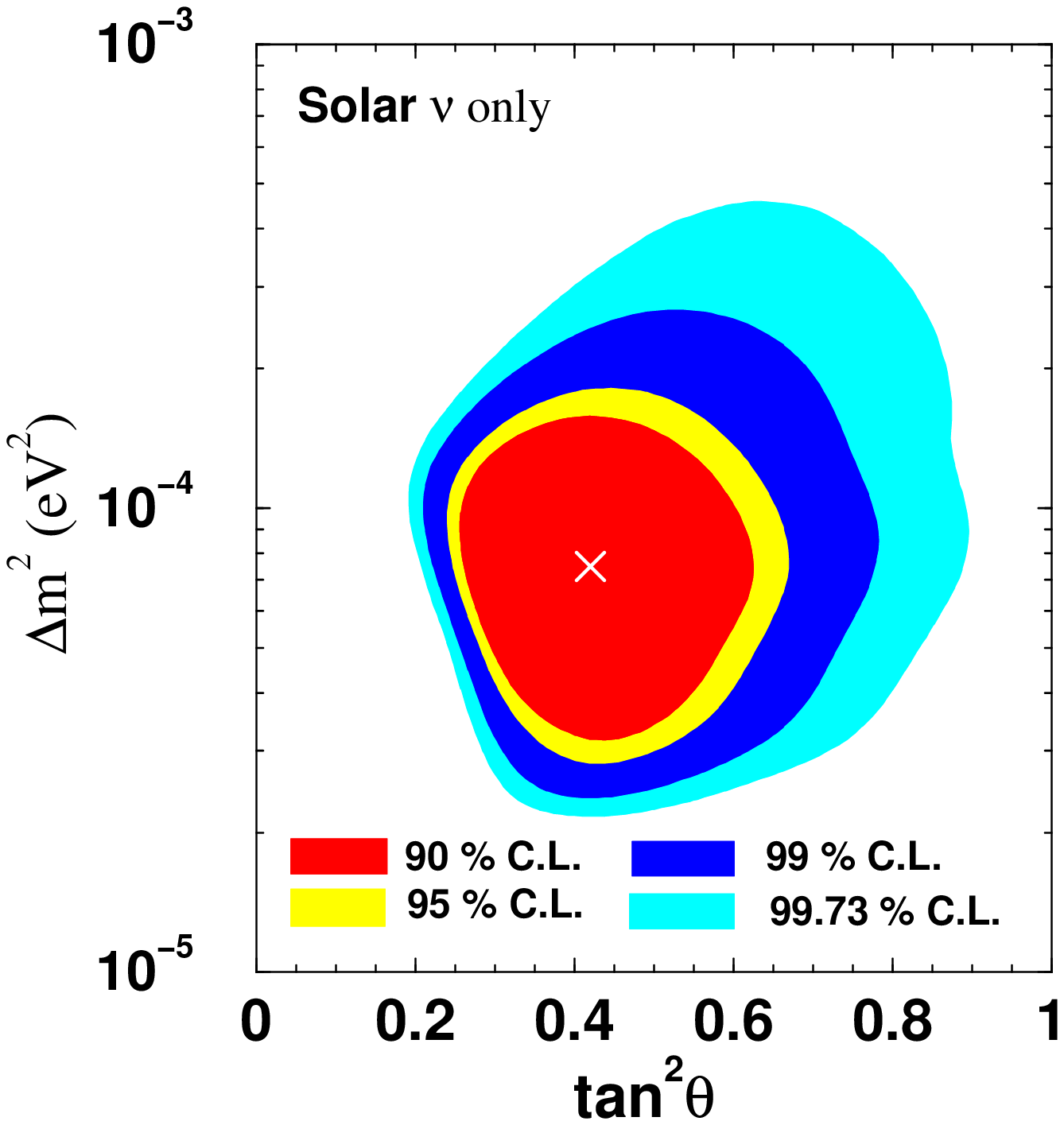} 
\vglue -.8cm
\caption{Region in $(\tan^2 \theta,\Delta m^2)$ plane 
allowed by the Super-Kamiokande-I zenith spectrum 
combined with rates from Homestake, GALLEX/GNO, SAGE and SNO.  
The best fit point is marked by a cross.}
\label{fig1} 
\centering\leavevmode 
\hglue -0.2cm
\vglue -.6cm
\includegraphics[scale=0.67]{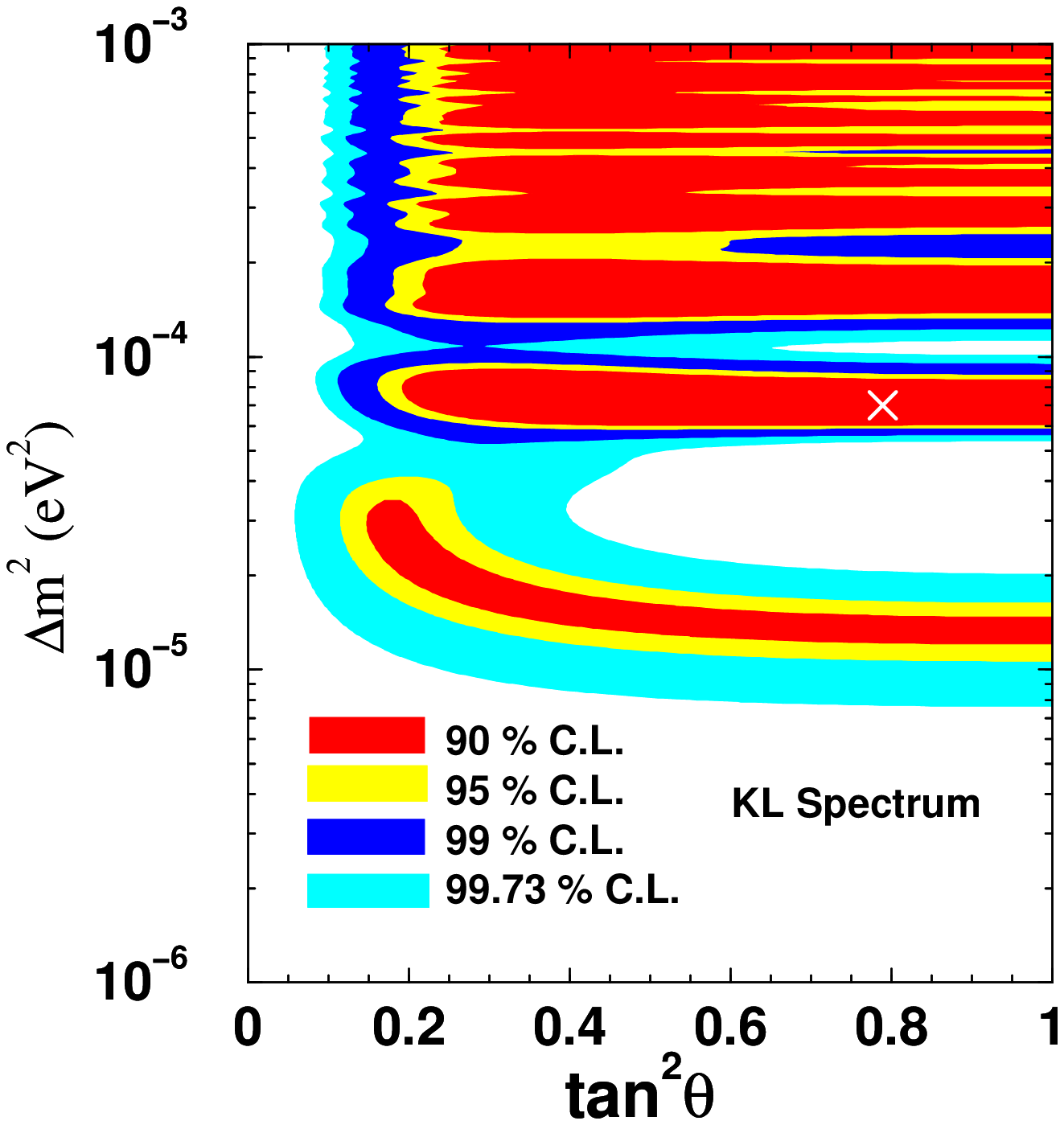} 
\vglue -0.7cm 
\caption{Regions in $(\tan^2 \theta,\Delta m^2)$ plane allowed by  
KamLAND data alone. The best fit point is marked by a cross.} 
\label{fig2} 
\vglue -0.5cm
\end{figure}

 In Figure~\ref{fig1} we show the region, in the  
 $(\tan^2 \theta,\Delta m^2)$ plane,  allowed 
 by the Super-Kamiokande-I zenith spectrum data as well as by 
 the rates of all other solar neutrino experiments 
 at 90\%, 95\%, 99\% and 99.73\% C.L. In our fit we obtained a     
 $\chi^2_{\odot}(\text{min})=37.7$ for 47 d.o.f  (83 \% C.L.),  
 corresponding to the global best fit values  
 $\Delta m^2= 7.5 \times 10^{-5}$ eV$^2$ and  $\tan^2 \theta=0.42$.

\subsection{KamLAND} 
\label{subsec:kam} 
\vglue -0.3cm
 KamLAND~\cite{kamland} is a reactor neutrino oscillation experiment 
 searching for  $\bar \nu_e$ oscillation from over 16 power reactors 
 in Japan and South Korea, mostly located at distances that 
 vary from 80 to 344 km from the Kamioka mine, allowing KamLAND to 
 probe the LMA MSW neutrino oscillation solution to 
 the solar neutrino problem.

 The KamLAND detector consists of about 1 kton of liquid scintillator 
 surrounded by photomultiplier tubes that register the arrival of 
 $\bar \nu_e$ through the inverse $\beta$-decay reaction 
 $\bar \nu_e + p \to e^+ + n$, by measuring $e^+$ and the 2.2 MeV $\gamma$-ray 
 from neutron capture of a proton in delayed coincidence. The $e^+$ 
 annihilate in the detector, producing the total visible energy 
 $E$  which is related to the incoming $\bar \nu_e$ energy, 
 $E_\nu$, as  $E=E_\nu-(m_n-m_p)+m_e$, where $m_n$,
 $m_p$ and $m_e$ are respectively, the neutron, proton and electron mass.

  After 145.1 days of data taking, which corresponds to 162 ton yr
  exposure, KamLAND has measured 54 inverse $\beta$-decay events, where
  87 were expected without neutrino conversion. These events are
  distributed in 13 bins of 0.425 MeV above the analysis threshold of
  2.6 MeV (applied to contain the background under about 1 event).

  We have theoretically computed the expected number of events in the 
  $i$-th bin, $N_i^{\text{theo}}$, as 
  \begin{equation}
  N_i^{\text{theo}} = \int dE_\nu \, \sigma(E_\nu) \sum_k \phi_k(E_\nu) 
  P_{\nu_e \to \nu_e} \int_i dE \,R(E,E^\prime),
  \end{equation}
  where $R(E,E^\prime)$ is the energy resolution 
  function, $E$ the observed and $E^\prime$ the 
  true $e^+$ energy, with the energy resolution $7.5\%/\sqrt{E(\text{MeV})}$. 
  Here $\sigma(E_\nu)$ is the neutrino interaction cross-section and $\phi_k$ 
  is the neutrino flux from the $k$-th power reactor, we have included all  
  reactors with baseline smaller than 350 km in the sum. 
  $P_{\nu_e \to \nu_e} \equiv P_{\bar \nu_e \to \bar \nu_e}$ 
 (if CPT is conserved, which we will assume here) is the familiar 
  neutrino survival probability in vacuum (the matter effect is negligible 
  here),  which is equal to one in case of no oscillation, 
  and explicitly depends on $\Delta m^2$ and  $\tan^2 \theta$.

  We were able to compute the region, in the $(\tan^2 \theta,\Delta m^2)$ 
  plane, allowed by the KamLAND spectrum data, by minimizing with respect to 
  these free parameters, the $\chi^2_{\text{KL}}$ function defined as 
  $\chi^2_{\text{KL}} = \chi^2_{\text{G}} + \chi^2_{\text{P}}$ with 
  \begin{equation}
  \chi^2_{\text{G}}= \displaystyle \sum_{i} 
  \frac{( N_i^{\text{theo}} - N_i^{\text{obs}})^2}{\sigma_i^2},
  \end{equation} 
  and 
  \begin{equation}
  \chi^2_{\text{P}}= \sum_{j} 2(N_j^{\text{theo}} - N_j^{\text{obs}}) + 
   2 \, N_j^{\text{obs}} \ln \displaystyle \frac{N_j^{\text{obs}}}
  {N_j^{\text{theo}}},
  \end{equation} 
  where  $\sigma_i = \sqrt{ N_i^{\text{obs}}+(0.0642\, N_i^{\text{obs}})^2}$ 
  is the statistical plus systematic uncertainty in the number of 
  events in the $i$-th bin and the sum in $i (j)$ is done over the bins 
  having 4 or more (less  than 4) events. 
  We have also computed the allowed regions using purely Gaussian or 
  Poissonian $\chi^2$ functions and found that the hybrid $\chi^2$ definition 
  above could reproduce better KamLAND's allowed regions~\cite{kamland}. 
  Therefore, we have prefered to use it in our paper 
  (see also Ref.~\cite{valle}). 

  Using this $\chi^2_{\text{KL}}$ we have computed the allowed region
  at 90\%, 95\%, 99\%  and 99.73\% C.L. shown in Fig.~\ref{fig2}, 
  which are quite consistent with the ones obtained by the KamLAND group  
  in Fig. 6 of Ref.~\cite{kamland}. 
  In our fit we obtained a  $\chi^2_{\text{KL}}(\text{min})=5.4$ for 
  11 d.o.f  (91 \% C.L.), corresponding to the best fit values  
  $\Delta m^2= 7.0 \times 10^{-5}$~eV$^2$ and  $\tan^2 \theta=0.79$.

\vglue -0.5cm
\subsection{Combined Results} 
\label{subsec:comb} 
\vglue -0.3cm 
 Combining the results of all solar experiments with KamLAND data 
 we have obtained the allowed region showed in Fig.~\ref{fig3}.
 The minimum value of $\chi^2_{\text{tot}}=\chi^2_\odot+\chi^2_{\text{KL}}$
 for the combined fit is      
 $\chi^2_{\text{tot}}(\text{min})=43.6$ for 60 d.o.f  (94.5 \% C.L.),  
 corresponding to the best fit values  
 $\Delta m^2= 7.1 \times 10^{-5}$~eV$^2$ and  $\tan^2 \theta=0.42$.
 We observe that there are two separated regions which are allowed  at  
 99 \% C.L.: a lower one in $\Delta m^2$ (region 1) where the global 
 best fit point is located, and an upper one (region 2) where the local best 
 fit values are $\Delta m^2= 1.5 \times 10^{-4}$~eV$^2$ 
 and  $\tan^2 \theta=0.41$, corresponding to 
 $\chi^2_{\text{loc}}(\text{min})=49.2$. We observe that 
 depending on the definition  of  $\chi^2_{\text{KL}}$ 
 (gaussian, poisson or hybrid) used, a 
 third tiny region above $\Delta m^2= 2 \times 10^{-4}$~eV$^2$ appears at 
 99.73\% C.L. However, apart from this small change, the combined 
 allowed region is not essentially affected by the $\chi^2_{\text{KL}}$ used.

 In Fig.~\ref{fig4} we show the theoretically predicted energy spectra 
 at KamLAND for no oscillation, the best fit values of the oscillation 
 parameters for KamLAND data alone and for KamLAND combined with solar 
 data in regions 1 and 2.
 We note that the fourth energy bin, which is for the moment below the 
 analysis cut, can be quite important in determining the values of 
 the oscillation parameters in the future.

\section{Future Perspectives} 
\label{sec:fut} 
\vglue -0.3cm  
 In this section we consider the effect of possible experimental 
 improvements which can help in determining the oscillation parameters 
 with more accuracy in the future.
 We first consider a reduction of the error in the SNO neutral-current 
 measurement then an increase of event statistics in KamLAND. 

\subsection{Effect of reducing SNO neutral-current error} 
\label{subsec:sno} 
\vglue -0.3cm  
 In order to constrain the solar neutrino oscillation parameters even more, 
 in particular, to decide in which of the 99\% C.L. islands $\Delta m^2$ 
 really lie, we have investigated the effect of increasing the SNO 
 neutral-current  data precision to twice its current value.  
 We have re-calculated the region, in the  $(\tan^2 \theta,\Delta m^2)$ plane, 
 allowed by all current solar neutrino data, artificially decreasing 
 the SNO NC measurement error but keeping the current central value, 
 as well as the other solar neutrino data, unchanged.
 The result can be seen in Fig.~\ref{fig5}. The best fit point 
 and the value of $\chi^2_\odot$(min) remain practically unchanged with 
 respect to the result obtained in Sec.~\ref{subsec:solar}, but the 
 allowed region shrinks significantly. This is because the $^8B$
 neutrino flux  normalization, which can be directly inferred from  
 SNO NC measurements, gets more constrained. 
 Combining this  with KamLAND data we obtain the allowed region 
 shown in Fig.~\ref{fig6}. We observe that this allowed region is 
 substantially smaller compared to the one shown in Fig.~\ref{fig3}. Moreover, 
 region 2 only remains at 99\%~C.L.

\subsection{Effect of increasing KamLAND statistics} 
\label{subsec:kamfut} 
\vglue -0.3cm  
  We simulate the expected KamLAND spectrum after one year of data taking 
  for three distinct assumptions. We have generated KamLAND future data 
  compatible with the best fitted values of $\Delta m^2$ and $\tan^2 \theta$ 
  obtained for : (a) KamLAND data alone,
 (b) KamLAND and current solar neutrino data in region 1 and 
 (c) KamLAND and current solar neutrino data in region 2.
  We have also included an extra bin, corresponding to the fourth bin in 
  Fig.~\ref{fig4}.  We have re-calculated the region allowed by the 
  combined fit with the current solar neutrino data in each case.
 
  The results of our calculations can be seen in Figs.~\ref{fig7}-\ref{fig9}.
  If the future KamLAND result is close to the current 
  one (see Fig.~\ref{fig7}), values 
  of $\tan^2 \theta$ larger than the ones allowed now will be possible and 
  region 2 will be excluded at 99\% C.L. For this case we have obtained 
  $\chi^2_{\text{tot}}$(min)~$=42.1$.
  On the other hand, if the future KamLAND data are more compatible with the 
  current best fit point of solar neutrino data (see Fig.~\ref{fig8}), 
  the global allowed region will diminish substantially with respect 
  to Fig.~\ref{fig3} and region 2 will only remain at 99\% C.L. 
  For this case we have obtained  $\chi^2_{\text{tot}}$(min)~$=39.1$.
  Finally, if after one year KamLAND data is more compatible with region 2 
  (see Fig.~\ref{fig9}) then one should observe an increase towards 
  larger values of $\Delta m^2$  in the combined allowed region with 
  respect to the one shown in Fig.~\ref{fig3}. In this case region 1 and 2 
  will have similar statistical significance, corresponding 
  to $\chi^2_{\text{tot}}$(min)~$ \sim 44.0$.

\section{Discussions and Conclusion} 
\label{sec:conclusions} 
\vglue -0.3cm

 We have performed a combined analysis of the complete set of solar neutrino 
 data with the recent KamLAND result in a two neutrino flavor oscillation 
 scheme. We have obtained, in agreement with  other groups~\cite{outros}, 
 two distinct islands, denominated regions 1 and 2, in the 
 $(\tan^2 \theta, \Delta m^2)$ plane, which are the most probable 
 regions where the true values of these  parameters lie. 
 Region 1, where the global best fit point was found, 
 is around $\Delta m^2=7.1 \times 10^{-5}$ eV$^2$, while region 2 
 is around $\Delta m^2=1.5 \times 10^{-4}$ eV$^2$.

 We have considered two possible future improvements in the determination 
 of the neutrino oscillation parameters.
 First, we have investigated the effect of a 50 \% decrease in 
 the error of the SNO NC measurement. We have shown that this would  
 substantially reduce the allowed parameter region when combined 
 with  KamLAND data. In particular, region 2 would not be allowed  at 
 95\% C.L. anymore.
 
\begin{figure} 
\centering\leavevmode 
\vglue 2.0cm 
\hglue -1.5cm
\includegraphics[scale=1.2]{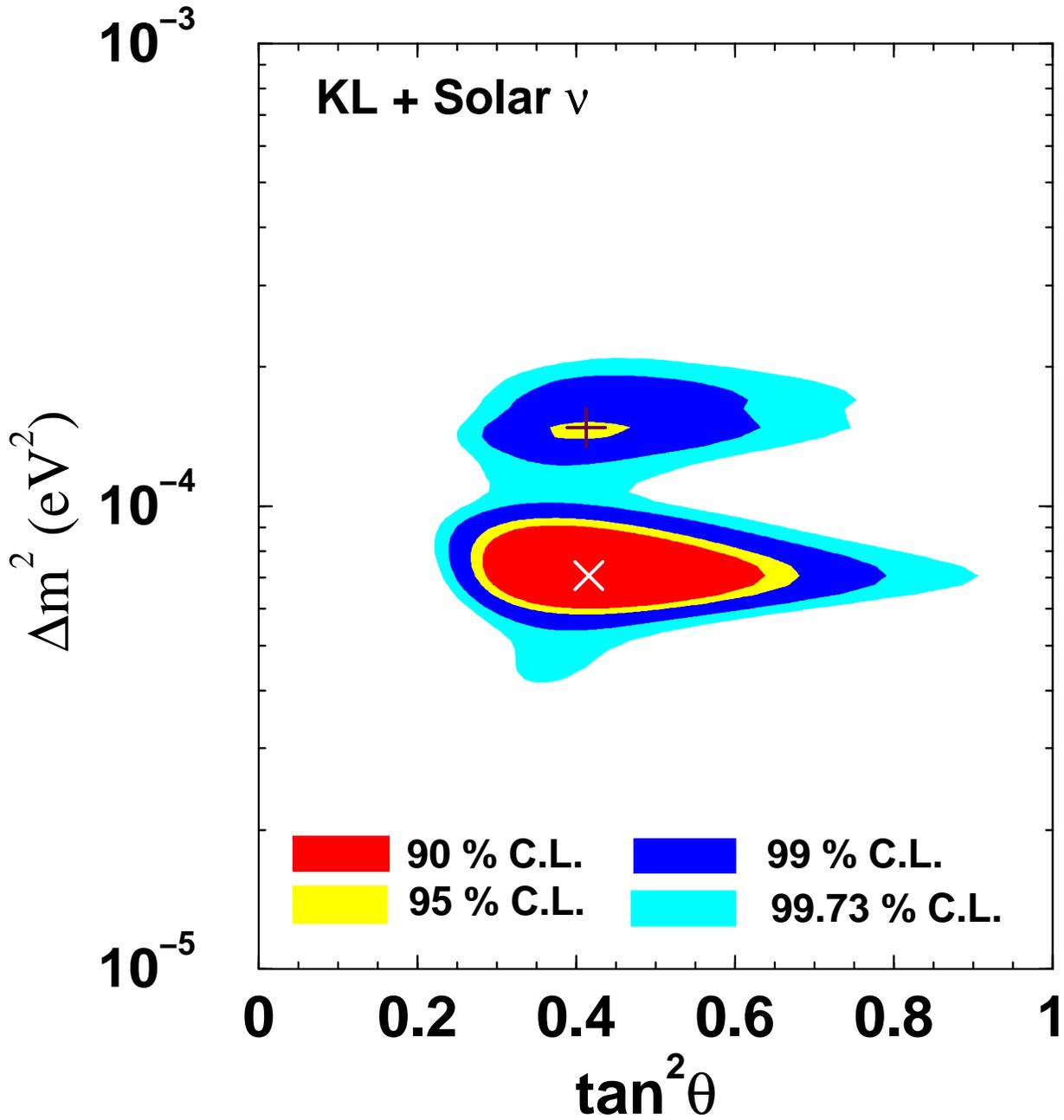} 
\vglue -.6cm 
\caption{Region allowed by all the solar neutrino experiments combined with 
KamLAND (KL) data. The region below (above) $\Delta m^2=10^{-4}$ eV$^2$ 
is referred to as region 1 (2).  The best fit points in each region 
are also marked by cross (global best) and plus (local best).} 
\label{fig3} 
\vglue -0.5cm
\end{figure}

\begin{figure} 
\centering\leavevmode 
\vglue -2.2cm  
\hglue -1.0cm
\includegraphics[scale=1.0]{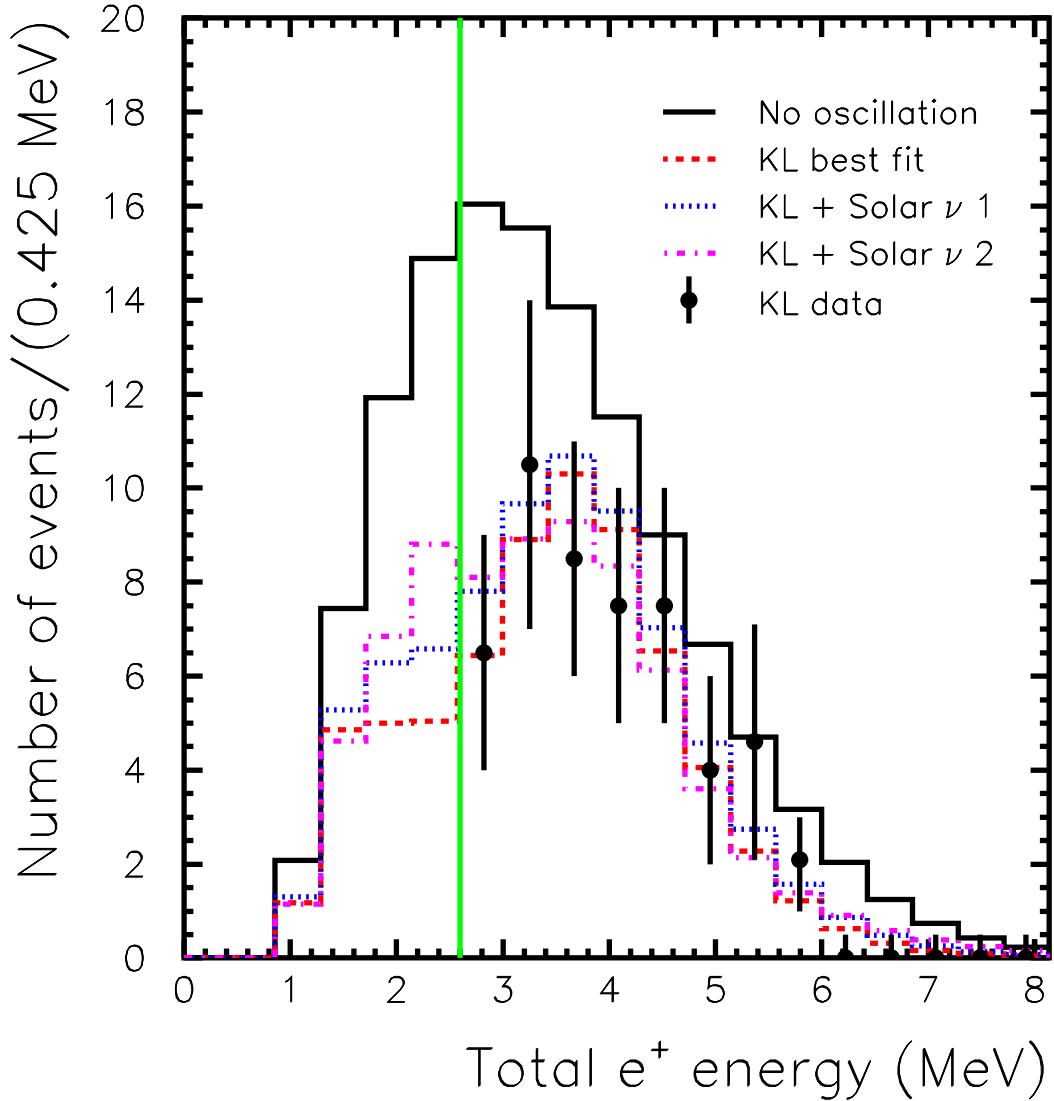} 
\vglue -6.cm 
\caption{Expected positron energy spectra at KamLAND (KL)
for no oscillation, the best fit values of the oscillation 
parameters for KamLAND data alone and KamLAND data combined with 
the solar neutrino data in regions 1 and 2 of Fig.~\ref{fig3}. 
The KamLAND data~\cite{kamland} is also shown as solid circles with error bars.
The energy threshold at 2.6 MeV is marked by a vertical line.} 
\label{fig4} 
\vglue 0.6cm
\end{figure}

\begin{figure} 
\centering\leavevmode 
\hglue -0.2cm
\vglue -.5cm 
\includegraphics[scale=0.7]{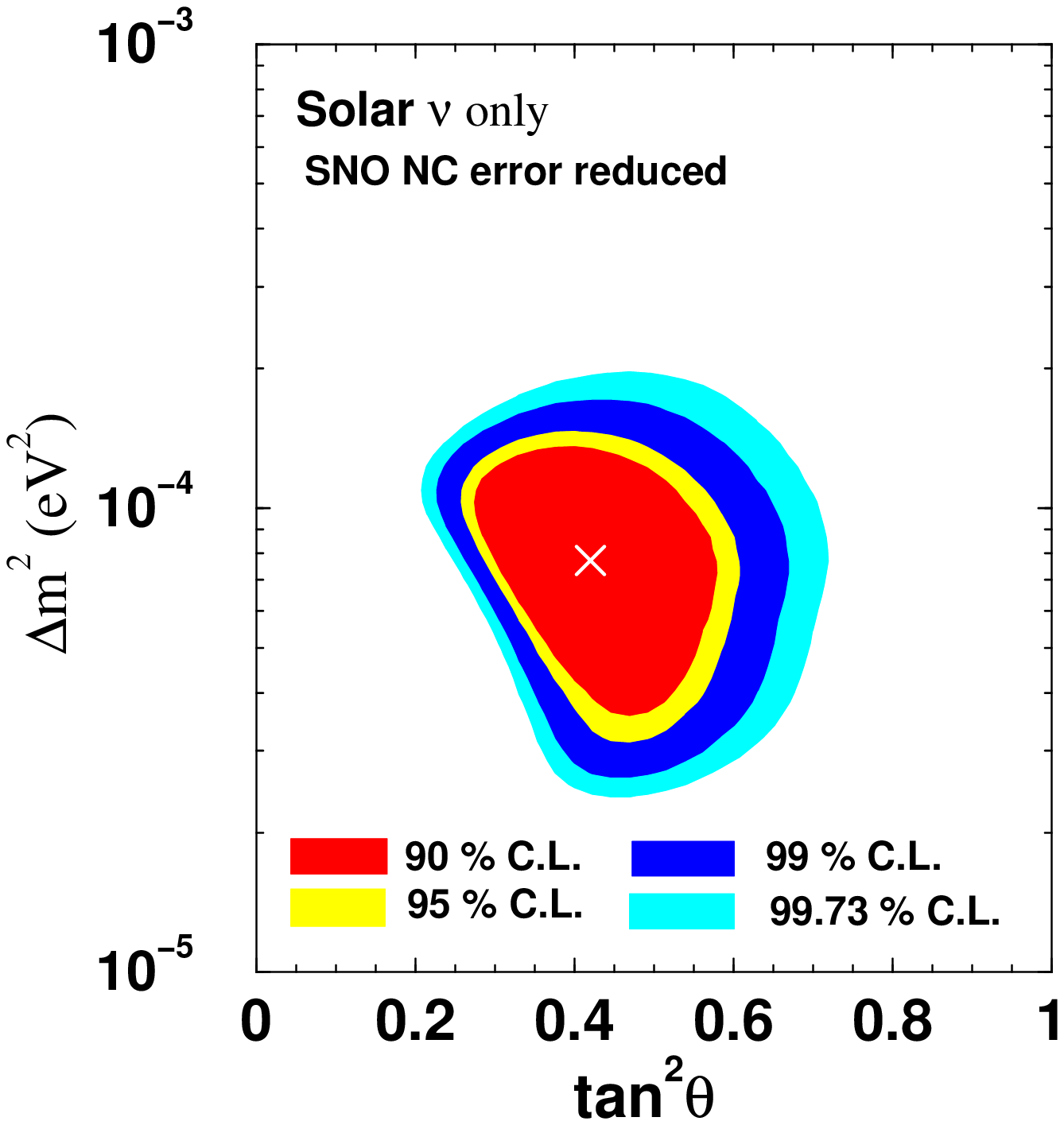} 
\vglue -.8cm 
\caption{Same as Fig.~\ref{fig1} but decreasing the SNO 
 neutral-current data error to half of its current value.}
\label{fig5} 
\centering\leavevmode 
\hglue -0.2cm
\vglue -0.2cm 
\includegraphics[scale=0.7]{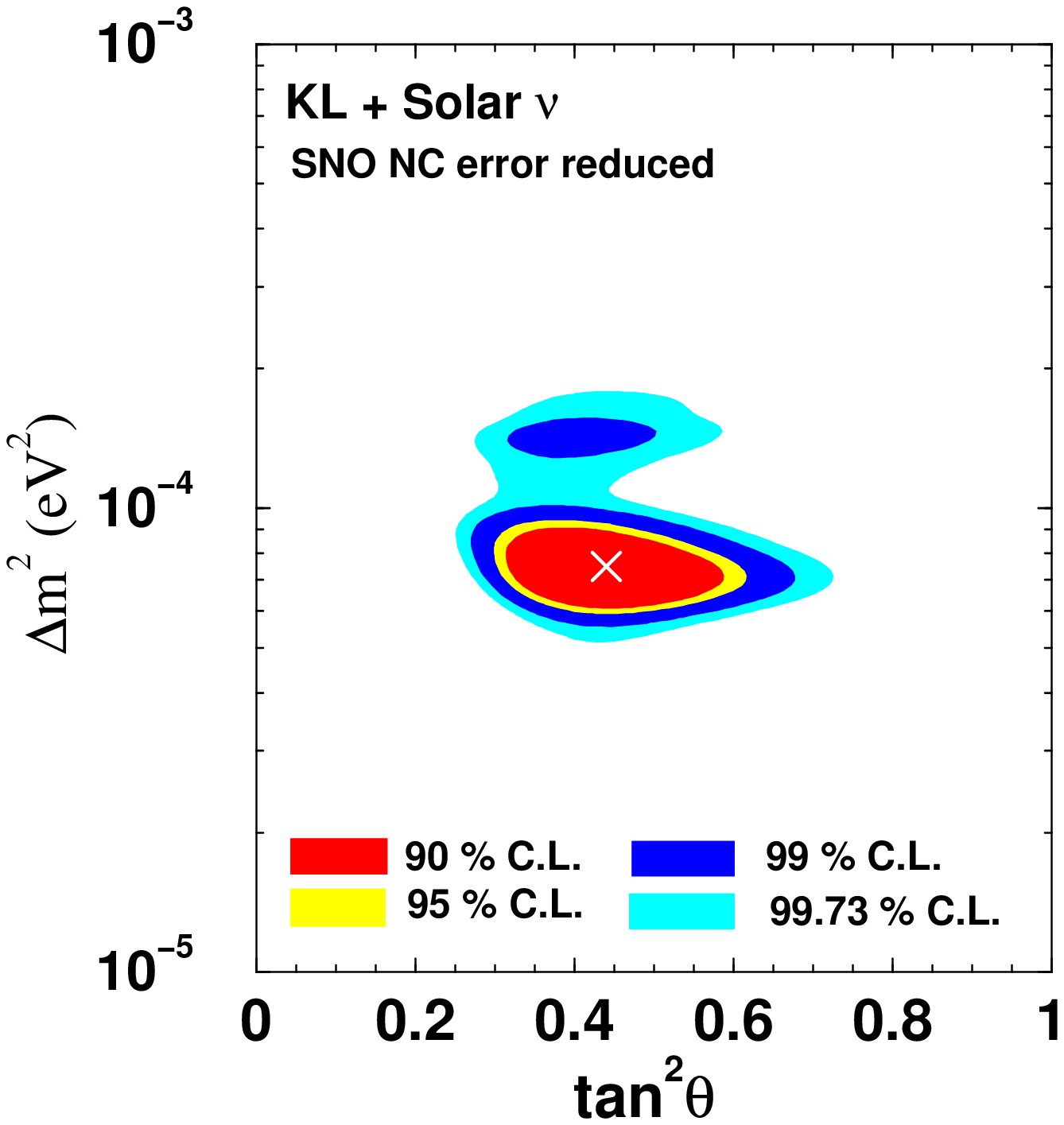} 
\vglue -0.8cm 
\caption{Same as Fig.~\ref{fig3} but decreasing the SNO 
 neutral-current data error to half of its current value.}
\label{fig6} 
\vglue -0.8cm 
\end{figure}

\begin{figure} 
\centering\leavevmode 
\hglue -0.2cm
\vglue -1.cm 
\includegraphics[scale=0.7]{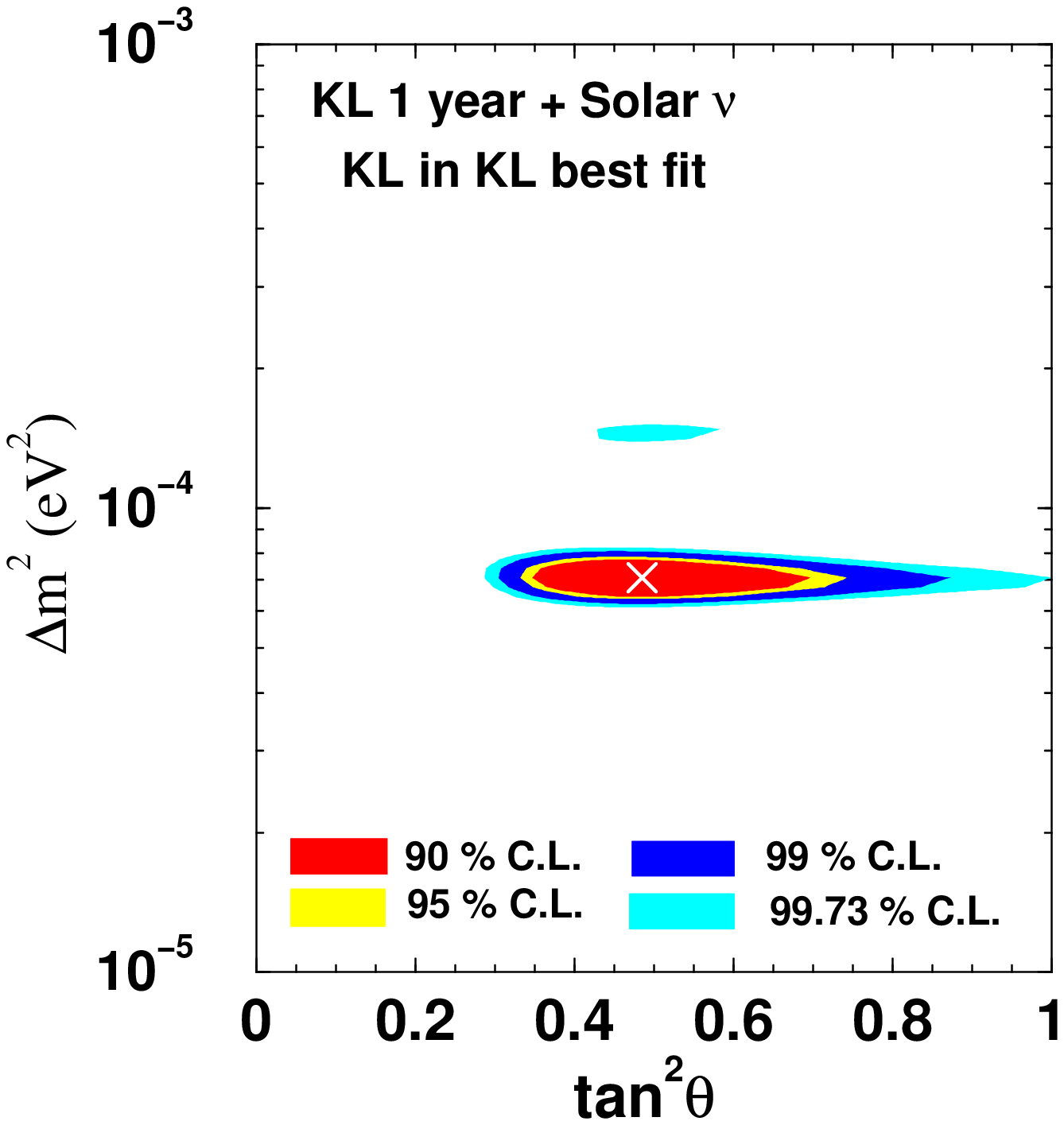} 
\vglue -0.6cm 
\caption{Same as Fig.~\ref{fig3} but for a simulated KamLAND spectrum 
after one year of data taking compatible with  the KamLAND alone best 
fit $\Delta m^2=7\times 10^{-5}$~eV$^2$ and $\tan^2\theta=0.79$.}
\label{fig7} 
\centering\leavevmode 
\hglue -0.2cm
\vglue -0.4cm 
\includegraphics[scale=0.7]{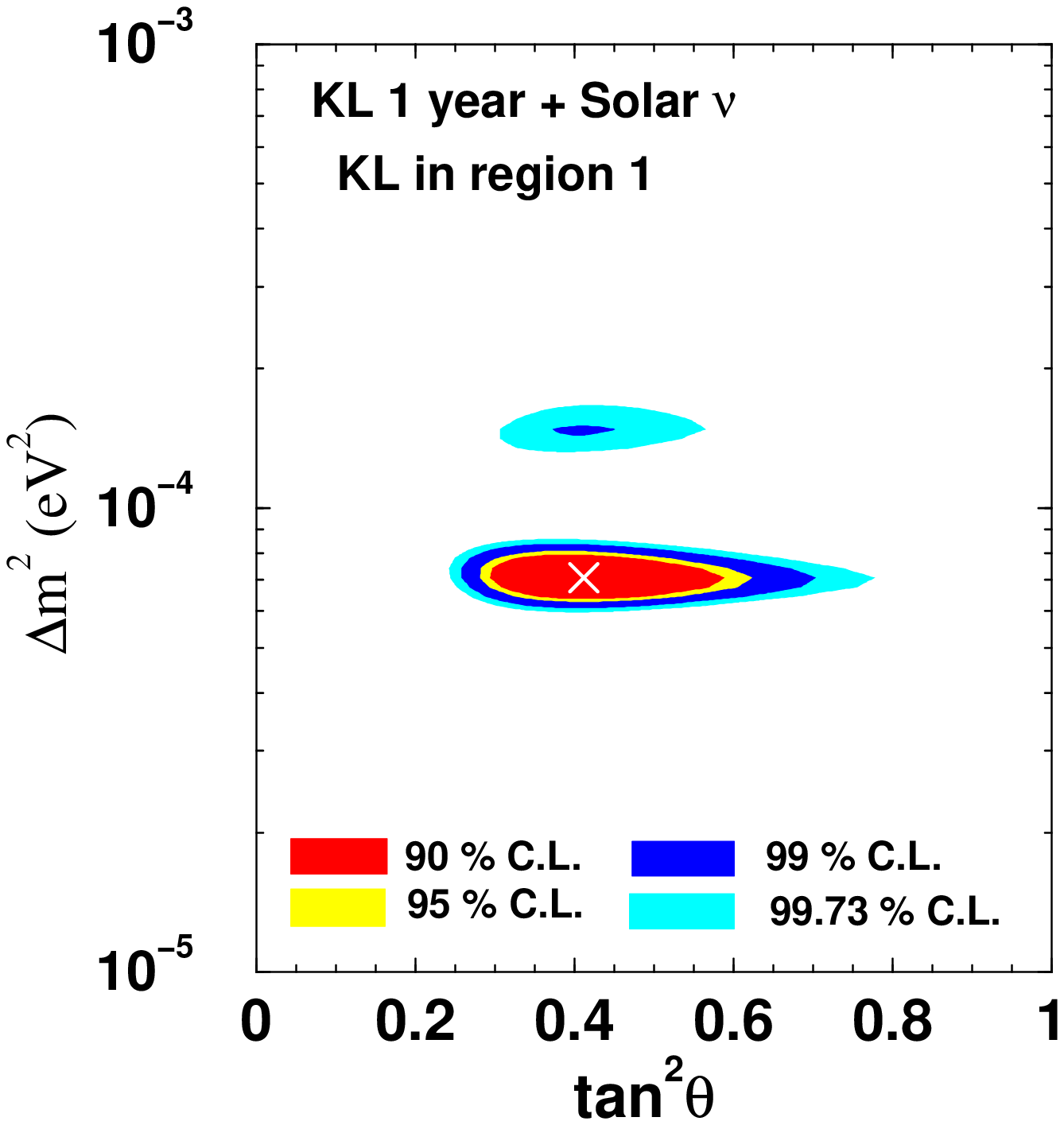}
\vglue -0.6cm 
\caption{Same as Fig.~\ref{fig7} but for the KL + Solar neutrino 
global best fit $\Delta m^2=7.1\times 10^{-5}$~eV$^2$ 
and $\tan^2\theta=0.42$ in region 1.}
\label{fig8} 
\vglue -0.3cm 
\end{figure}

\begin{figure} 
\centering\leavevmode 
\hglue -0.2cm
\vglue 0.cm 
\includegraphics[scale=0.7]{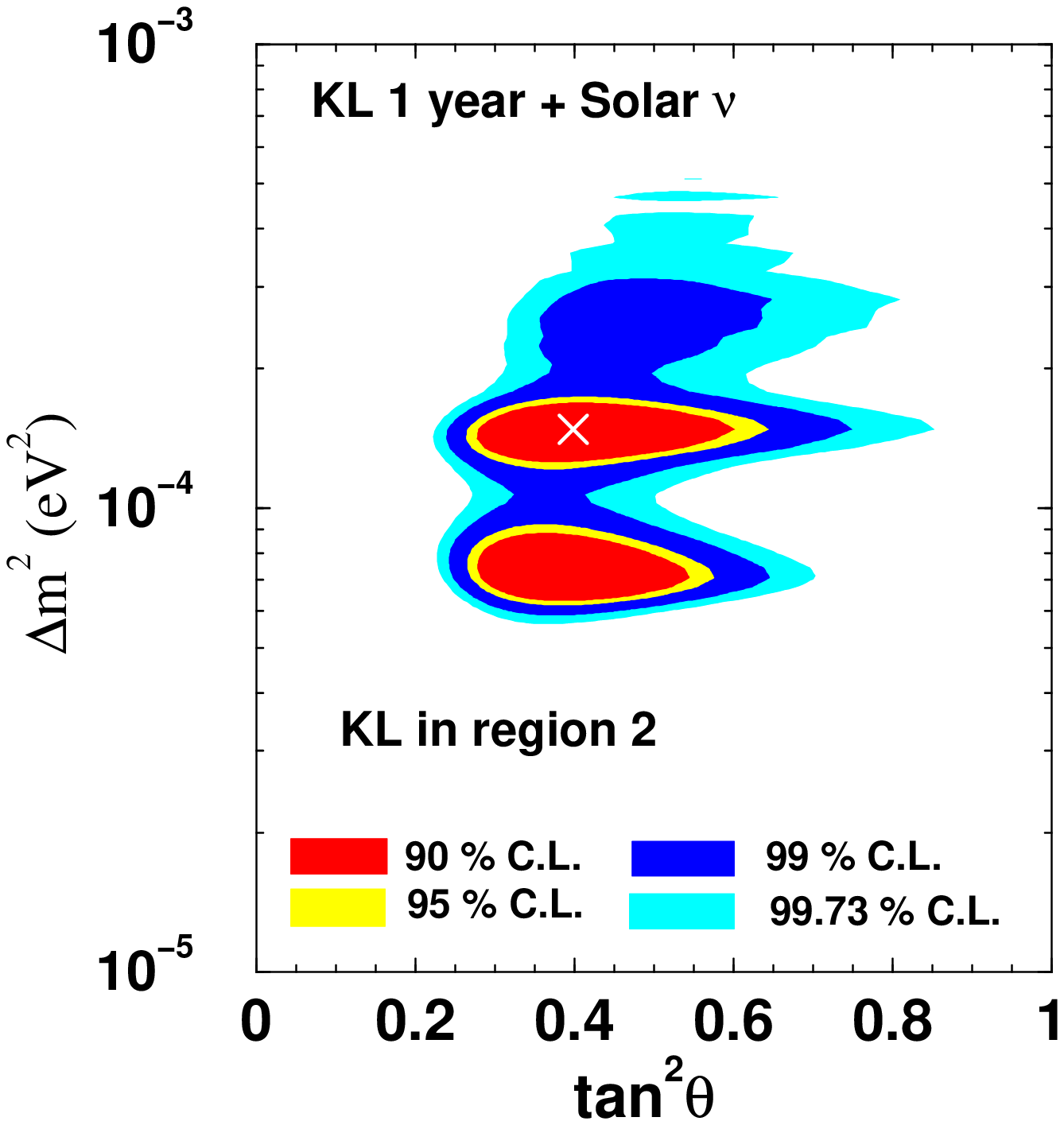}
\vglue -0.6cm 
\caption{Same as Fig.~\ref{fig7} but for the KL + Solar neutrino 
local best fit $\Delta m^2=1.5\times 10^{-4}$~eV$^2$ and $\tan^2\theta=0.41$
in region 2.}
\label{fig9} 
\vglue -0.3cm 
\end{figure}

 Second, we have studied what can happen in the near future, when 
 KamLAND collects 1 year of data. We have simulated the expected KamLAND 
 spectrum including an extra lower bin, corresponding to the fourth bin in 
 Fig.~\ref{fig3}. Three different cases were studied in combination with 
 the present solar neutrino data. In the first case, we have assumed that 
 the future KamLAND  spectrum will be compatible with oscillation 
 parameter values at the best fit point for the present KamLAND data 
 alone. This is the most restrictive case for region 2. 
 In the second case, we have considered that future data 
 will be more compatible with the present best fit point for the solar 
 neutrino experiments. In this case, the combined allowed region will be 
 much smaller than the present one and region 2 will be only allowed at 
 99\%~C.L. Finally, in the third case, we have assumed that the future 
 KamLAND data will be compatible with the local best fit point in region 2.
 In this case, the combined allowed region will suffer an increase towards 
 larger values of $\Delta m^2$  and region 1 and 2 will both have  
 similar statistical significance.

\begin{acknowledgments} 
 This work was supported by Funda{\c c}{\~a}o de Amparo 
 {\`a} Pesquisa do Estado de S{\~a}o Paulo (FAPESP) and Conselho 
 Nacional de  Ci{\^e}ncia e Tecnologia (CNPq).
\end{acknowledgments} 
 

 \vglue -0.5cm 

 

\begin{thebibliography}{99} 

\bibitem{atmnuobs}  
Y. Fukuda {\em et al.} (Super-Kamiokande Collaboration),  
Phys. Rev. Lett. {\bf 81}, 1562 (1998); 
%
H. S. Hirata {\em et al.} (Kamiokande Collaboration),  
Phys. Lett. B {\bf 205}, 416 (1988);  
{\it ibid.}\ {\bf 280}, 146 (1992);  
Y. Fukuda {\em et al.}, {\it ibid.}\  {\bf 335}, 237 (1994);  
%
R. Becker-Szendy {\em et al.} (IMB Collaboration),  
Phys. Rev. D {\bf 46}, 3720 (1992);  
%
M. Ambrosio {\em et al.} (MACRO Collaboration),  
Phys. Lett. B {\bf 478}, 5 (2000);  
B. C. Barish,  Nucl. Phys. B (Proc. Suppl.) {\bf 91}, 141 (2001);  
%
W. W. M. Allison {\em et al.} (Soudan-2 Collaboration),  
Phys. Lett. B {\bf 391}, 491 (1997);  
Phys. Lett. B {\bf 449}, 137 (1999);  
W. A. Mann, Nucl. Phys. B (Proc. Suppl.) {\bf 91}, 134 (2001). 


\bibitem{k2k} 
K2K Collaboration, S.~H.~Ahn {\it et al.}, 
Phys.\ Lett.\ B {\bf 511}, 178 (2001); 
K.~Nishikawa,  Talk presented at XXth International Conference  
on Neutrino Physics and Astrophysics (Neutrino 2002),  
May 25-30, 2002, Munich, Germany. 

\bibitem{kamland} KamLAND Collaboration, K.~Eguchi {\it et al.}, 
Phys.\ Rev.\ Lett.\  {\bf 90}, 021802 (2003);
see also http://www.awa.tohoku.ac.jp/KamLAND/index.html.

\bibitem{fits} 
J.~N.~Bahcall, M.~C.~Gonzalez-Garcia and C.~Pe\~na-Garay, 
JHEP {\bf 0207}, 054 (2002);
A.~Bandyopadhyay {\it et al.}, 
Phys.\ Lett.\ B {\bf 540}, 14 (2002);
V.~Barger {\it et al.}, Phys.\ Lett.\ B {\bf 537}, 179 (2002);
P.~Aliani  {\it et al.}, AIP Conf.\ Proc.\  {\bf 655}, 103 (2003)
[arXiv:hep-ph/0211062];
P.~C.~de Holanda and A.~Yu.~Smirnov, 
Phys.\ Rev.\ D {\bf 66}, 113005 (2002);
P.~Creminelli, G.~Signorelli and A.~Strumia, 
JHEP {\bf 0105}, 052 (2001);
G.~L.~Fogli  {\it et al.}, Phys.\ Rev.\ D {\bf 66}, 053010 (2002);
M.~Maltoni {\it et al.}, Phys.\ Rev.\ D {\bf 67}, 013011 (2003). 

\bibitem{exotics}
A.~M.~Gago {\it et al.}, 
Phys.\ Rev.\ D {\bf 65}, 073012 (2002);
M.~Guzzo  {\it et al.}, 
Nucl.\ Phys.\ B {\bf 629}, 479 (2002), and references therein.


\bibitem{homestake} 
Homestake Collaboration, B.~T.~Cleveland {\it et al.}, 
Astrophys.\ J.\ {\bf 496}, 505 (1998).

\bibitem{gallex}
GALLEX Collaboration, W.~Hampel {\it et al.},
Phys.\ Lett.\ B {\bf 447}, 127 (1999).

\bibitem{gno}
GNO Collaboration, M.~Altmann {\it et al.},
Phys.\ Lett.\ B {\bf 490}, 16 (2000);
T.~Kirsten, on behalf of GNO Collaboration, talk presented at
{\it Neutrino 2002}, May 25-30, Munich, Germany, 
see http://neutrino2002.ph.tum.de.


\bibitem{sage}
SAGE Collaboration, D.~N.~Abdurashitov {\it et al.}, 
Nucl.\ Phys.\ (Proc.\ Suppl.\ )  {\bf 91}, 36 (2001); 
Latest results from SAGE homepage: 
{\tt http://EWIServer.npl.washington.edu/SAGE/. }

\bibitem{superk} 
Super-Kamiokande Collaboration, S.~Fukuda {\it et al.},
Phys.\ Lett.\ B {\bf 539}, 179 (2002).
 
\bibitem{sno} 
SNO Collaboration, Q.\,R.\, Ahmad {\it et al.},
\prl{87}, 071301 (2001);
{\it ibid.} {\bf 89}, 011301 (2002); 
{\bf 89}, 011302 (2002).

\bibitem{msw} 
S.P. Mikheyev and A. Yu. Smirnov,  
Yad. Fiz. {\bf 42}, 1441 (1985) 
[Sov. J. Nucl. Phys. {\bf 42}, 913 (1985)];  
L. Wolfenstein, Phys. Rev. D {\bf 17}, 2369 (1978).

\bibitem{bahcall} 
J.~N.~Bahcall, {\it Neutrino Astrophysics}, Cambridge University Press, 
Cambridge, England, 1989.

\bibitem{mswtriangle} 
S.~J.~Parke, Phys. Rev. Lett. {\bf 57}, 1275 (1986);
S.~J.~Parke and T.~P.~Walker, Phys. Rev. Lett. {\bf 57}, 2322 (1986); 
J.~Bouchez {\it et al}, Z. Phys. {\bf C 32} (1986) 499;
M. Cribier {\it et al.}, Phys. Lett. {\bf B182}, 89 (1986);
S. P. Mikheyev and A. Yu. Smirnov,  
Proc. of the 12th Int. Conf Neutrino'86 (Sendai, Japan) 
eds. T. Kitagaki and H. Yuta, 177 (1986) and   
Proc. of the Int. Symp. on Weak and Electromagnetic Interactions in Nuclei, 
WEIN-86, (Heidelberg, 1986), 710.
 
\bibitem{ssm} 
J.~N.~Bahcall, M.~H.~Pinsonneault and S.~Basu, Astrophys.~J.
{\bf 555}, 990 (2001).

\bibitem{valle} M.~Maltoni {\it et al.} in Ref.~\cite{outros}.

\bibitem{outros} 
V.~Barger and D.~Marfatia, 
Phys.\ Lett.\ B {\bf 555}, 144 (2003),
G.~L.~Fogli, {\it et al.}, arXiv:hep-ph/0212127; 
M.~Maltoni, T.~Schwetz and J.~W.~F.~Valle, arXiv:hep-ph/0212129;
A.~Bandyopadhyay, {\it et al.}, arXiv:hep-ph/0212146;
J.~N.~Bahcall, M.~C.~Gonzalez-Garcia and C.~Pe\~na-Garay, 
JHEP {\bf 0302}, 009 (2003).

\end{thebibliography}
\end{document}